\documentclass[twocolumn,showpacs,showkeys,preprintnumbers]{revtex4}

\def\be{\begin{equation}}
\def\ee{\end{equation}}
\def\beq{\begin{eqnarray}}
\def\eeq{\end{eqnarray}}
\def\n{\nonumber}
\def\bay{\begin{array}}
\def\eay{\end{array}}

\begin{document}

\preprint{CIRI/02-swrskg01}
\title{What is the spacetime of {\em physically realizable\/} spherical collapse?}

\author{Sanjay M. Wagh$^1$, Ravindra V. Saraykar$^{1,2}$, and Keshlan S. Govinder$^3$}
\affiliation{
$^{1}$Central India Research Institute, Post Box 606, Laxminagar, Nagpur 440 022, India\\
E-mail:cirinag@nagpur.dot.net.in\\
$^{2}$Also at: Department of
Mathematics, Nagpur University Campus, Nagpur 440 010, India. \\
\phantom{m} \qquad E-mail: sarayaka@nagpur.dot.net.in\\ $^3$School
of Mathematical \& Statistical Sciences, University of Natal,
Durban 4041, South Africa.
\\E-mail: govinder@nu.ac.za}


\date{August 31, 2002}
\begin{abstract}
We argue that a particular spacetime, a spherically symmetric
spacetime with hyper-surface orthogonal, radial, homothetic
Killing vector, is a physically meaningful spacetime that
describes the problem of spherical gravitational collapse in its
full ``physical" generality.
\end{abstract}

 \maketitle

\newpage
\section{Introduction} \label{intro}
Any sufficiently sparsely distributed ordinary ``neutral" matter
is dusty, that is, collision-less and pressureless. Further,
emission of radiation and, hence, radiation is not expected in
such dust.

In Newtonian gravity, as well as in general relativity, we can
then study gravitational collapse from this dusty ``initial" state
of matter.

Under the action of its self-gravity, dust matter collapses. Then,
self-gravity leads to mass or energy-flux in some preferential
direction, the radial direction for a spherical spacetime.

But, this is not the flux of radiation. Therefore, there is no
mass-flux in the rest frame of collapsing dusty matter, but it is
present for other observers in the spacetime.

Next stage of collapse is reached when particles of dust begin to
collide with each other. Negligible amount of radiation, but
existing nonetheless, is expected from whatever atomic excitations
or from whatever free electrons get created in atomic collisions
in such matter. Therefore, dusty matter evolves into matter with
pressure and radiation, both {\em simultaneously} non-vanishing.

The energy-flux can no longer be removed by going to the rest
frame of matter.

As far as Newtonian gravity and general relativity, both, are
concerned, we may study the gravitational collapse beginning even
as matter with non-vanishing pressure and radiation.

At some stage, exothermic, thermonuclear reactions begin in matter
with non-negligible pressure. With it, a star is born in the
spacetime.

This is the manner of gravitational collapse of dusty matter
leading to the birth of a star. Till the exothermic thermonuclear
reactions in the stellar core support the overlying stellar
layers, such a stellar object is gravitationally stable.

But, the spacetime continues to be dynamic  since radiation is
present in it. The stellar object may also accrete matter from its
surrounding while emitting radiation.

Once again, in Newtonian gravity and in general relativity, both,
we may study the collapse of this ``initial" stellar configuration
of matter.

Now, as and when ``heating" of the overlying stellar layers
decreases due to changes in exothermic thermonuclear processes in
the core of the star, the self-gravity of the stellar object leads
to its gravitational contraction. These are, in general, very slow
and involved processes.

Gravitational contraction leads to generation of pressure by
compression and by the occurrence of exothermic thermonuclear
reactions of heavier nuclei. The star may stabilize once more.

This chain, of gravitational contraction of star, followed by
pressure increase, followed by subsequent stellar stabilization,
continues as long as thermonuclear processes produce enough heat
to support the overlying stellar layers.

The theory of the atomic nucleus shows that {\em exothermic\/}
nuclear processes do not occur when Iron nucleus forms. With time,
the rate of heat generation in iron-dominated-core becomes
insufficient to support the overlying stellar layers which may
then bounce off the iron-core resulting into a stellar explosion,
a supernova.

Then, many, different such, stages of evolution are the results of
physical processes that are unrelated to the phenomenon of
gravitation. These are, for example, collisions of particles of
matter, electromagnetic and other forces between atomic or
sub-atomic constituents of matter etc.

As an example, let some non-gravitational process, opposing
collapse, result into pressure that does not appreciably rise in
response to small contraction of the stellar matter. That is,
pressure does not appreciably rise when gravitational field is
increased by a small amount Then, the collapse of a sufficiently
massive object would not be halted by that particular
non-gravitational process. Therefore, a {\em mass limit\/} is
obtained in this situation. For example, electron degeneracy
pressure leads to the Chandrasekhar limit \cite{stars}.

Clearly, some of the non-gravitational processes determine the
gravitational stability of physical objects. This is true in
Newtonian gravity as well as in general relativity, both.

In general relativity, non-gravitational processes are included
via the energy-momentum tensor for matter. Non-gravitational
processes determine the relation of density and pressure of
matter. The temporal evolution of matter is to be determined from
such a relation, and from other physical relations, if any.

It is therefore that, by {\em physically realizable gravitational
collapse}, we mean collapse that leads matter, step by step,
through the above different ``physical" stages of evolution.

Hence, the spacetime of ``physically realizable" collapse of
matter must be able to begin with any stage in the chain of
evolution of matter under the action of its self-gravity. The
temporal evolution from any ``initial" data, any ``physical" stage
in question, is to be obtained from applicable non-gravitational
properties of matter.

But, a supernova remnant, or a star that failed to explode, may be
quite massive for its self-gravity to dominate over all
conceivable competing reasons opposing it at various stages of
further evolution. This may also happen as a result of
mass-accretion taking the object in question over some mass-limit
in operation. The collapse is, now, {\em unstoppable}. A spacetime
singularity is expected to form in such unstoppable collapse.

Associated with studies of unstoppable collapse is the issue of
whether the {\em physically realizable\/} gravitational collapse
leads to a black hole or to a naked spacetime singularity. This is
the issue of the Cosmic Censorship Hypothesis (CCH)
\cite{penrose98}.

Clearly, the answer to this very important question in general
relativity can then be obtained only on the basis of the spacetime
of the physically realizable gravitational collapse.

Now, we show below that, for spherical symmetry, a 
spacetime with hyper-surface orthogonal, radial, homothetic
Killing vector provides ``all" the above steps of evolution  of
matter.

\section{Spacetime of physically realizable collapse} One
radially homothetic spacetime has the following metric \cite{cqg1}
in co-moving coordinates: \be
ds^2=-y^2dt^2+\gamma^2(y')^2B^2dr^2+y^2Y^2d\Omega^2
\label{ssmetfinal}\ee with $y=y(r)$, an overhead prime indicating
a derivative with respect to $r$, $B\equiv B(t)$, $Y\equiv Y(t)$
and $\gamma$ being a constant.

As can be easily verified, the metric (\ref{ssmetfinal}) admits a
spacelike Homothetic Killing Vector (HKV) of the form \be
X^a\;=\;(0,\frac{y}{\gamma y'},0,0) \label{hkvradial} \ee This is
therefore the case of hyper-surface orthogonal, spacelike HKV.

Now, the spacetime of (\ref{ssmetfinal}) is required, by
definition, to be locally flat at all of its points including the
center. The condition for elementary flatness at the center of
(\ref{ssmetfinal}) is \be {y'|}_{r\,\sim\,0}\;\approx\;1/\gamma
\label{conrzero} \ee This condition must be imposed on any $y(r)$.
With this condition, (\ref{conrzero}), the HKV of metric
(\ref{ssmetfinal}) is, at the center,
$y|_{r=0}\,\partial/\partial_r$.

Now, we may use the function $y(r)$ in (\ref{ssmetfinal}) as a new
radial coordinate - the area coordinate - as long as $y' \neq 0$.
However, the situation of $y'=0$ represents a coordinate
singularity that is similar to, for example, the one on the
surface of a unit sphere where the analogue of $y$ is
$\sin{\theta}$ \cite{synge}. We can change the radial coordinate
to suitable one before such a coordinate singularity is reached.

\subsection*{Singularities and degeneracies of metric (\ref{ssmetfinal})}
The Ricci scalar for (\ref{ssmetfinal}) is: \beq {\cal R} &=&
\frac{4\dot{Y}\dot{B}}{y^2YB}+\frac{2\ddot{B}}{y^2B}
-\frac{6}{y^2\gamma^2B^2}\n \\
&&\qquad\qquad  +\frac{2}{y^2Y^2} +\frac{2\dot{Y}^2}{y^2Y^2}
+\frac{4\ddot{Y}}{y^2Y} \eeq

Then, there are two types of genuine curvature singularities of
(\ref{ssmetfinal}), namely, the first type for vanishing of
temporal functions for some $t=t_s$ and, the second type for
$y(r)=0$ for some $r$.

The ``physical'' distance corresponding to the ``coordinate''
radial distance $\delta r$ is $\ell\;=\;\gamma (y') B \delta r$.
Then, collapsing matter forms the spacetime singularity in
(\ref{ssmetfinal}) when $\ell=0$, {\em ie}, $B(t)\,=\,0$ for it at
some $t\,=\,t_s$. Thus, the singularity of first type is a
singular hyper-surface for (\ref{ssmetfinal}).

The singularity of the second type is a singular sphere of
coordinate radius $r$. The singular sphere reduces to a singular
point for $r=0$ that is the center of symmetry. Singularities of
the second type constitute a part of the initial data, singular
data, for the evolution.

The metric (\ref{ssmetfinal}) has evident degeneracies when
$y(r)=0$, $y(r)\to \infty$ either on a degenerate sphere of
coordinate radius $r$, for some ``thick" shell or globally.
Another degeneracy occurs for $y(r)=constant$ for some ``thick"
shell or globally.

In what follows, we shall assume that there is no singular
initial-data and that there are no evidently degenerate situations
for the metric (\ref{ssmetfinal}).

\section*{Temporal evolution in (\ref{ssmetfinal})}
The Einstein tensor for (\ref{ssmetfinal}) is: \beq G_{tt}&=&
\frac{1}{Y^2}-\frac{1}{\gamma^2B^2} + \frac{\dot{Y}^2}{Y^2} +
2\frac{\dot{B}\dot{Y}}{BY}
\\ G_{rr}&=& \frac{\gamma^2B^2y'^2}{y^2} \left[-\,2\frac{\ddot{Y}}{Y}
-\frac{\dot{Y}^2}{Y} \right. \n \\ && \qquad \qquad \qquad \left.
+\frac{3}{\gamma^2B^2} - \frac{1}{Y^2}\right]
\\G_{\theta\theta}&=&-\,Y\,\ddot{Y}-Y^2\frac{\ddot{B}}{B}
- Y\,\frac{\dot{Y}\dot{B}}{B}+\frac{Y^2}{\gamma^2B^2}
\\G_{\phi\phi}&=& \sin^2{\theta}\,G_{\theta\theta} \\
G_{tr}&=&2\frac{\dot{B}y'}{By}  \label{gtr} \eeq

Clearly, the energy-flux depends on $\dot{B}$. Now, define the
quantity \be \sigma \equiv \,{{\sigma}^{1}}_{1} =
{{\sigma}^{2}}_{2} = - \frac{1}{2}{{\sigma}^{3}}_{3} =
\frac{1}{3y}\left( \frac{\dot{Y}}{Y}-\frac{\dot{B}}{B} \right)
\label{shearscalar} \ee Here, $\sigma_{ab}$ represents the
shear-tensor of the fluid and the shear-scalar is given by
$\sqrt{6}\;\sigma$. Therefore, the spacetime of (\ref{ssmetfinal})
is, in general, shearing and radiating, both.

With our assumptions of no singular and degenerate initial data,
we then have a ``cosmological" situation  - continued spherical
collapse of matter from the assumed ``initial" state.

Now, for the co-moving observer with four-velocity $ {\bf
U}\,=\,\frac{1}{y}\;\frac{\partial}{\partial t}$, the {\em
radial\/} velocity of the fluid is $ V^r =\dot{Y}$ where an
overhead dot denotes a time derivative. The co-moving observer is
accelerating for (\ref{ssmetfinal}) since
$\dot{U}_a={U_a}_{;\,b}U^b$ is, in general, non-vanishing for
$y'\neq 0$. The expansion is $ \Theta =
\frac{1}{y}\,\left(\,\frac{\dot{B}}{B} \;+\;
2\,\frac{\dot{Y}}{Y}\,\right) $ Now, we turn to steps of collapse
of matter as outlined in \S \ref{intro}.

\subsection*{Step I - Evolution of dust}
Consider the collapse from ``dusty" stage without radiation. Then,
for vanishing energy-flux, $B\approx {\rm constant}\equiv B_o$.

Then, the co-moving density, $\rho$, of dust is \be \rho
\,=\,\frac{1}{y^2}\,\left[ \frac{\dot{Y}^2}{Y^2} \, +
\frac{1}{Y^2}-\frac{1}{\gamma^2B_o^2} \right]
\label{dustdensity}\ee and the function $Y(t)$ is determined by
the condition of vanishing of the isotropic pressure: \be
4Y\ddot{Y}+\dot{Y}^2+1-\zeta\,Y^2 = 0 \ee Here, $\zeta =
5/\gamma^2B_o^2$, a positive constant.

A solution of this equation is obtainable as \be \frac{d
Y}{\sqrt{-1+\zeta/5 Y^2 + c_o Y^{-1/2}}} = t - t_0\ee where $c_o$
is constant. Since $\dot{Y}$ is the radial velocity of matter for
the co-moving observer, we require that solution for which
$\dot{Y}\to 0$ for $t\to -\infty$.

\subsection*{Step II - Evolution with pressure and radiation}
Now, pressure and radiation, both, get {\em simultaneously\/}
switched on in the spacetime of (\ref{ssmetfinal}) when $B(t) \neq
0$. This is as per the expectation that dusty matter evolves to
one with simultaneous occurrence of pressure and radiation, both.

Now, the co-moving density is \be \rho =\frac{1}{y^2}\,\left[
\frac{\dot{Y}^2}{Y^2} \,+\,2\,\frac{\dot{Y}\dot{B}}{YB} +
\frac{1}{Y^2}-\frac{1}{\gamma^2B^2} \right] \label{density} \ee

We also obtain \be 2\,\frac{\ddot{Y}}{Y} +
\frac{\ddot{B}}{B}\;=\;\frac{2}{\gamma^2B^2}-\frac{y^2}{2}(\rho +
3p) \label{rddot} \ee

Then, from (\ref{rddot}), the relation of pressure and density of
matter is the required additional ``physical" information. Also
required is other relevant ``physical" information to determine
the radiation generation in the spacetime of (\ref{ssmetfinal}).

To provide for the required information of ``physical" nature is a
non-trivial task in general relativity just as it is for Newtonian
gravity. The details of these considerations are, of course,
beyond the scope of this letter.

However, it is clear that the field equations determine only the
temporal functions from the properties of matter in the spacetime
of (\ref{ssmetfinal}).

Moreover, it is also clear that matter will continue to pile up on
such a star in a ``cosmological setting" and, hence, such a star
will always be taken over any mass-limit in operation at any stage
of its evolution that will, ultimately, lead to the singular
hyper-surface of the spacetime of (\ref{ssmetfinal}).

The radial dependence of matter properties is ``specified" as
$1/y^2$ but the field equations of general relativity do not
determine the metric function $y(r)$ in (\ref{ssmetfinal}).

Therefore, the radial distribution of matter is {\em arbitrary} in
terms of the co-moving radial coordinate $r$. This is the
``maximal" {\em physical\/} freedom compatible with the assumption
of spherical symmetry, we may note. Note, however, that the
physical generality here is not be taken to mean the
``geometrical" generality.

\section{Issue of regularity of center}
A spherical spacetime admits an $SO(3)$ group of rotational
symmetry. The orbits of the symmetry group are closed ones. The
{\em center\/} of the spherically symmetric spacetime geometry is
defined to be the ``invariant point" of the $SO(3)$ group of
rotations, as it must be.

Galilean invariance of the Newtonian equations implies that {\em
every\/} observer observes the shrinkage of orbits of the rotation
group to zero radius at the center of a spherically symmetric
object.

Consequently, we may demand that the orbits of the rotation group
also shrink to zero radius for a spherically symmetric spacetime.
Such a spacetime is said to possess a {\em regular\/} center.

Now, for (\ref{ssmetfinal}), $y(r)$ is the ``area radius". When
$y|_{r=0} \neq 0$, the orbits of the rotation group $SO(3)$ do not
shrink to zero radius at the center of the spacetime although the
curvature invariants remain finite at the center. Also, when
$y|_{r=0} = 0$, the orbits of the rotation group shrink to zero
radius at the center but the curvature invariants blow up at the
center, then.

It is well-known \cite{mcintosh} that the center and the initial
data for matter, both, are not {\em simultaneously\/} regular for
a spherical spacetime with hyper-surface orthogonal HKV.

Therefore, the spacetime of (\ref{ssmetfinal}) does not possess a
{\em regular\/} center and {\em regular\/} matter data,
simultaneously, and we may consider it to be ``unphysical" even
when its matter follows the expected ``physical" evolution.

However, this issue requires careful analysis. It is crucial to
ask: who, which observer, is observing the orbits of the rotation
group shrink to zero radius? This is important in general
relativity though not in Newtonian gravity.

Consider the Schwarzschild spacetime. The asymptotic observer does
not see any sphere, centered on the mass-point, shrink to any
radius below $r=2M$, the ``infinite red-shift surface".

Then, for the asymptotic observer, $r=2M$ is the center of the
spacetime! Thus, the ``area radius" has this minimum value for the
asymptotic observer.

It is no coincidence that this minimum value of the ``area radius"
as observed by the asymptotic observer is related to the
gravitational mass or the Schwarzschild mass $M$.

Recall the well-know Newtonian theorem: ``The gravitational force
on a body that lies outside a closed spherical shell of matter is
the same as it would be if all the shell's matter were
concentrated into a point at its center."

The general relativistic manifestation of this Newtonian result is
that the spacetime of a spherical body must possess non-vanishing
central value for mass in that spacetime. For suitable observer,
the orbits of the rotation group are not expected to shrink to
zero radius at the center of a spherical spacetime with matter.

The co-moving observer of (\ref{ssmetfinal}) is the ``equivalent"
of the asymptotic observer of the Schwarzschild spacetime in that
it is also the ``cosmological" observer for (\ref{ssmetfinal}).
Then, for this observer, orbits of the rotation group $SO(3)$ are
not expected to shrink to zero ``area radius" at the center of the
spacetime of (\ref{ssmetfinal}).

Therefore, the conflict of the ``non-regularity" of the center and
the ``physical" evolution of matter in (\ref{ssmetfinal}) can be
resolved with this observation.

Now, let us call a star for which the orbits of the rotation group
do not shrink to zero radius at its center a {\em strange\/} star,
in contrast to the ``standard" spherical star for which the center
is regular. Then, it is clear from the above that {\em all\/} the
spherical stars embedded in a cosmological surrounding are
expected to be {\em strange\/} stars in the sense described above.

\section{Concluding remarks}
Many spacetimes of spherically symmetric nature are known
\cite{exactsoln}. For example, the ``original" Schwarzschild
\cite{abrams}, the ``standard" Schwarzschild (or, Hilbert-Droste,
\cite{abrams}), the Vaidya, the Tolman-Bondi class, the
Friedmann-Lemaitre-Robertson-Walker spacetimes. Some of these
known examples of spherically symmetric spacetime geometries
contain no matter, only radiation dust, only matter dust, etc.

To construct the spacetime of a ``physically realizable" spherical
collapse of matter, we therefore need to ``match" different,
appropriate, such spacetime geometries.  Of course, this is a
non-trivial and, mostly, very difficult task.

But, these different spacetime geometries are, clearly, different
choices of $y(r)$ in (\ref{ssmetfinal}). For example, consider
collapse from initial dusty matter with vacuum ``exterior". Then,
$y(r)$ is infinite in the exterior.

The spacetime of (\ref{ssmetfinal}) then provides the ``final"
spacetime that may be obtained after matching many different
spacetimes with appropriate physical conditions of matter. It is,
therefore, the spacetime of {\em physically realizable collapse\/}
of spherically symmetric matter.

We have, therefore, studied the shear-free collapse in
\cite{sfcollapse} and the collapse with shear and energy flux in
\cite{sphcollapse}. We have also obtained source free
electromagnetic fields using the Hertz-Debye formalism
\cite{cohen} in \cite{sphcollapse}. In these studies, some general
conclusions as well as the explicit forms for temporal metric
functions have been presented for a simple equation of state of
the barotropic form $p=\alpha\rho$ where $\alpha$ is a constant.
We have also studied, in \cite{hawking}, the phenomenon of Hawking
radiation in a radially homothetic spacetime of the metric
(\ref{ssmetfinal}).


\end{document}